\documentstyle[aasms4,12pt,epsf]{article}
\def\sun{\odot}
\def\lsim{\, \lower2truept\hbox{${< \atop\hbox{\raise4truept\hbox{$\sim$}}}$}\,}
\def\gsim{\, \lower2truept\hbox{${> \atop\hbox{\raise4truept\hbox{$\sim$}}}$}\,}

\begin{document}

\title{Starburst in the Intragroup Medium of Stephan's Quintet\footnote{Based on observations made with ISO, an ESA project with
instruments
funded by ESA Member States and with the participation of ISAS and NASA.}}
\author{Cong Xu}
\affil{Infrared Processing and Analysis Center, Jet Propulsion Laboratory, 
Caltech 100-22, Pasadena, CA 91125}
\author{Richard Tuffs}
\affil{Max-Planck-Institut f\"ur Kernphysik,
Postfach 103980, D69117 Heidelberg, Germany}
\received{July 24, 1998}
\accepted{August 25, 1998}

\begin{abstract}
Based on new
ISO mid-infrared observations and ground based $H_\alpha$ and
near-infrared observations, we report the detection of a bright starburst
in the intragroup medium (IGM) of the famous compact group of galaxies
Stephan's Quintet (Source A in Fig.1). We demonstrate that
this starburst is caused by a collision between a high 
velocity ($\delta$V$\sim$ 1000 km/sec) intruder galaxy (NGC7318b) 
and the IGM of
the group. While this is the only starburst known today that is induced
by a galaxy/cold-intergalactic-medium collision, it provides new
constraints to the theory 
for interaction-induced starbursts, and may hint at a
new mechanism for the star formation excess seen in more distant
clusters.
\end{abstract}


\section{Introduction}
Starbursts (Rieke et al. 1980; Gehrz et al 1983) are ``cosmic fireworks" where
hundreds of thousands of massive stars that are up to a million times
brighter than the Sun are born and die in a short time scale of a few
10$^7$ years (compared to the age of most galaxies of $\sim 10^{10}$
years).  Despite the wide interest, many basic questions about
starbursts remain unanswered.  First of all, how are they triggered?
Theoretically 
the only well established triggering mechanism for starbursts involves
low velocity encounters/mergers 
between galaxies (Larson  \& Tinsley 1978; Joseph et al. 1984; Lonsdale
et al. 1984; Noguchi  \& Ishibashi 1986; Mihos \& Hernquist 1996).
In this letter, we shall show that a bright starburst
in the intragroup medium (IGM) of the famous compact group of galaxies
Stephan's Quintet is triggered by a new mechanism, namely  
a collision between a high velocity ($\delta$V$\sim$ 1000 km/sec)
intruder galaxy (NGC7318b) and the IGM of
the group.

Stephan's Quintet (hereafter SQ), 
discovered 120 years ago (Stephan 1877),
is perhaps the most famous example of the compact
galaxy group class that is characterized by aggregates  of 4-8 galaxies
with implied space densities as high as those in cluster
cores (Hickson 1982).
SQ has been observed with almost every large
telescope and in almost every domain of the electromagnetic spectrum.
These data suggest (Moles et al. 1997)
that SQ has been visited by at least
two ``intruders'' (NGC7320c and NGC7318b) in the past few 10$^8$
years.  Consequently, its intragroup medium (IGM) is filled with HI
gas stripped from its spiral members (Shostak et al. 1984).
In an attempt to study the star formation activity and 
the dust properties of this IGM we made observations at
11.4$\mu m$, 15$\mu m$ (ISOCAM), 60$\mu m$ and 100$\mu m$ (ISOPHT)
using the Infrared Space Observatory (Kessler et al. 1996).
In these observations, an outstanding bright source 
(Source A in Fig.1) is detected in
a region quite far away from centers of member galaxies.
In this letter, we publish the ISOCAM
15$\mu m$ data (the rest of ISO data will be published in subsequent papers) 
which, together with data from 
the following-up gound based H$_\alpha$ and NIR observations,
demonstrate conclusively that this bright source is associated with
a starburst in the IGM of SQ.

\section{Observations}
\subsection{ISOCAM observations}
The ISOCAM observations at 15$\mu m$ (Fig.1) were made on
1996 May 23 using the ISOCAM Long-Wavelength-Channel array (32$\times 32$
pixels). Raster scans with $PFOV=6''$,
$M=12$, $\delta M=48''$ (in-scan) and $N=20$, $\delta N=6''$
(cross-scan) were made in the micro-scan mode (CAM01) using filter LW3
($\lambda_0=15\mu m$, $\delta\lambda =6\mu m$). 
The 1$\sigma$ rms noise measured in the inner part ($\sim
5'\times 3'$) of the image is 0.04 mJy/pixel, and is  
higher near the edges of the
image. The absolute calibration is taken
from the ISOCAM Observer's Manual (1994), which has an uncertainty of
less than 30\% (Sauvage, private communication). The angular resolution
(FWHM) of the image is 10$''$. The basic data reduction was done using 
the CAM Interactive Analysis (CIA)
software\footnote{CIA is a joint development by ESA Astrophysics
Division and the ISOCAM Consortium led by the ISOCAM PI, C. Cesarsky,
Direction des Sciences de la Matiere, C.E.A., France.}.

\subsection{H$_\alpha$ observations}
The H$_\alpha$ observations (Fig.2) were made
in August 1997 using the 2.2m telescope of the
observatory of Max-Planck-Institut at Calar Alto (Spain).
In order to separate the H$_\alpha$ emission associated with the
intruder galaxy NGC7318b ($v\simeq 5700$ km/sec) 
from the H$_\alpha$ emission associated with
the rest of the group ($v\simeq 6600$ km/sec),
CCD images of the H$_\alpha$--[NII] emission
were obtained using two narrow band 
filters 667/7 and 674/8, centered at 6667{\AA} and 6737{\AA}
with FWHM bandwidths of 66{\AA} and 76{\AA}, respectively.
The transmission of the 667/7 filter for the H$_\alpha$ line
of redshift 0.019 ($v=5700$ km/sec) is 0.49,
and for the H$_\alpha$ line of redshift 
0.022 (6600 km/sec) is 0.04.
The transmissions of the 674/8 filter for the same two
components of the H$_\alpha$ emission are 0.06 and 0.49,
respectively. The 5700 km/sec component of the
H$_\alpha$--[NII] emission is therefore estimated from the
image obtained using the 667/7 filter, and the 
6600 km/sec component from the 674/8 image. 
The continuum in both images is subtracted using an R-band CCD image
of the same field.
To enhance the sensitivity for the diffuse emission,
both H$_\alpha$--[NII] maps (original 
resolution $\sim 1''$) are smoothed to 
a round Gaussian beam with FWHM=2$''$. Note that 
in some regions the two components have similar morphology.
This could be an indication that the emission in those regions is
rather associated with another component, namely the one with
recession velocity of 6000 km/sec (Moles et al. 1997;
Shostack et al. 1984). However this uncertainty does
not affect seriously the H$_\alpha$ fluxes (which are
sums of the 5700 km/sec and the 6600km/sec components) 
reported in Table 1, because while the 6000km/sec component
might have been counted twice, its flux would have been underestimated 
by about
a factor of two in both the 667/7 map and the 674/8 map,
given that the transmissions of the
two filters for the 6000 km/sec component are only about half of the 
transmissions for the 5700 km/sec and the 6600 
km/sec component, respectively. 

\subsection{Near Infrared K'-band observation}

A near infrared K'-band ($\lambda_0=2.1\mu m$)
image of the central part of Stephan's Quintet
were obtained using the 3.5m telescope of the observatory of
Max-Planck-Institut at Calar Alto (Spain), mounted with
the 256$\times 256$ pixels (pixel size =$0.''81$) MAGIC camera.
The observations were made in March 1997. The weather conditions were
photometric and stellar photometry consistent to $\sim 2\%$ throughout
the night. The seeing was about $1''$. The image covers 
about $4' \times 4'$ sky (Fig.2). In the central $3'\times 3'$ field,
the 1$\sigma$ rms noise is 20.4 mag/arcsec$^2$. Because of the mosaic,
the noise is higher near the edges.

\section{Results}

\subsection{A starburst (Source A) far away from galaxy centers}
Figure 1 presents the central $\sim 4'\times 4'$ region of
the 15$\mu m$ ISOCAM image of Stephan's Quintet
(false color).  As shown by Sauvage et
al. (1996),
the 15$\mu m$ emission is a good, nearly extinction
free star formation rate indicator. Located slightly
upper-right from the map center, Source A lies about
$1'$ ($\sim 25$ kpc at $D=80$ Mpc where H$_0$=75) north of NGC7318b and
about $1'$ west of NGC7319.  It covers a region of $\sim$ 30$''$ and
its peak surface brightness in the 15$\mu m$ map is  second only to
that of the Sy2 nucleus of NGC7319. 
At 80Mpc the 15$\mu m$ flux density of Source A corresponds to a 
luminosity ($\nu L_\nu$) $L_{15\mu}= 4.8\; 10^8 L_\odot$,
about 30 times higher than the $L_{15\mu}$ ($1.7\; 10^7 L_\odot$) 
of
the foreground Sd galaxy NGC7320 (assumed to be at 10Mpc; Table 1).

In an early photographic observation including both the $H_\alpha$
emission and the underlying continuum emission, Arp (1973)
identified several HII regions in the region of Source A.
H$_\alpha$ emission was also detected in the Source A region   
in a recent published H$_\alpha$ image centered on the Sy2 galaxy
NGC7319 (Moles et al. 1997)
though Source A falls on the edge of that image
and therefore is not fully covered.
A grid photometry map for SQ by Schombert et al. (1990)
reveals a concentration of very
blue data points (B$-$V= 0.3--0.5) in this region. The
H$_\alpha$ knots and blue color provide strong support for an
interpretation that the MIR emission is due to dust heated by massive
young stars. 

In order to firmly establish the starburst nature of Source A,
we carried out new H$_\alpha$ and K$'$ (2.1$\mu m$) 
imaging observations using an optical CCD and a near-infrared
detector array, respectively (Section 2.2 and 2.3). 
With two narrow band filters (667/7 and 674/8),
the 5700 km/sec component and the 
6600 km/sec component of the H$_\alpha$ emission, the
former being associated with the current intruder galaxy NGC7318b and the
latter with the rest of the group (Table 1), are separated.
In Fig.2 the contours of the two H$_\alpha$
components are overlaid on the K$'$ image. Indeed strong
H$_\alpha$ emission from both components is detected in the Source A
region, and the position of the H$_\alpha$ peak (associated with the
6600km/sec component) coincides well with that of the
15$\mu m$ emission peak.
The H$_\alpha$ equivalent line width, EW(H$_\alpha$), of the region is
115$\pm 15${\AA}. Such high EW(H$_\alpha$) is found only in 
starbursts (Keel et al. 1985).

The star formation rate (SFR) in the Source A region can be estimated from
both the 15$\mu m$ and the H$_\alpha$ luminosities.
A ratio between flux densities of the $15\mu m$ and of the
thermal radio at 6cm f$_{15\mu}$/f$_{th,6cm}=85$ (mJy/mJy) is
assumed, taken from the median value of the star formation regions in
M51 (Fig.2b of Sauvage et al. (1996)).
Following Kennicutt et al. (1994)
and adopting a Salpeter IMF with $m_l=0.1$ M$_\sun$ and
$m_u=100$ M$_\sun$, we obtain from Source A's 15$\mu m$ luminosity
(4.8 10$^8$ L$_\sun$) an SFR of
0.81 $M_\sun/yr$. Assuming the same IMF, the H$_\alpha$ luminosity
(2.2 10$^7$ L$_\sun$)
gives an SFR of 0.66 $M_\sun/yr$. 
The two estimates of the SFR agree with each other within the
uncertainties. This also suggests that the extinction of
the H$_\alpha$ emission in Source A is low.

For starbursts, 
according to Bruzual  \& Charlot (1993),
the stellar-mass to NIR luminosity ratio is 
insensitive to the starburst age, and therefore the NIR luminosity
is a good stellar mass indicator.
Using the model of Bruzual  \& Charlot (1993) and assuming that the
stellar population in Source A is the product of a starburst of age 
10$^7$ --- 10$^8$ years (the widely accepted 
age span of starbursts), the stellar
mass in Source A can be estimated from its
K$'$ magnitude (Table 1), which is in the range 
0.8 --- 1.6 10$^7$ M$_\sun$.
 Dividing this mass by the star formation
rate derived above, one finds that the starburst in Source A is 
only $\sim 1$---2 10$^7$ years old. 
Any possible contribution from an underlying older population
(e.g. from stripped stars)
may only reduce the estimate of this age, because less NIR light
would then be due to stars formed in the current burst.

\subsection{Triggering mechanism: IGM--intruder collision} 
How is the starburst in Source A, which is more than 20 kpc away from
centers of nearby galaxies, triggered? 
HI observations of SQ (Shostak et al. 1984)
reveal three velocity components
associated with the group (not including the foreground galaxy NGC7320): 
1) a 6600km/sec component
distributed throughout the intragroup space, 
2) a 6000km/sec component north of N7318b and 3) a
5700km/sec component located south of  N7318b. Moles et
al. (1997)
suggested that both the two latter components are
associated with NGC7318b, the velocity difference between them due to
the rotation.  Both the 6600 and 6000 km/sec components show local maxima
in the region of Source A, both with HI surface density of $\sim 3.3\;
10^{20}$ atoms/cm$^2$. The X-ray (Pietsch et al. 1997)
and the radio continuum (van der Hulst  \& Rots 1981) 
observations show a shock front, generated
by the collision between NGC7318b and the IGM, on the eastern side
of NGC7318b (also visible on our H$_\alpha$ maps, Fig.2). Noticeably,
Source A is located in a region {\it without} detected
X-ray emission.  ROSAT maps  (Pietsch et al. 1997)
show that Source A lies either: 1) in a gap in the shock
related X-ray emission (if the detached emission NW of Source A is
interpreted as an extension of the shock front) 
or 2) on the N edge of the shock front.

The HI, X-ray and radio continuum 
data in the literature and our MIR, H$_\alpha$ and NIR
data are all directing to a scenario such as the following:
During the intrusion of NGC7318b into SQ, 
a high speed ($\sim 1000$ km/sec)
collision between the local maxima of the two HI components in Source
A region, one (6600
km/sec component) associated with the intragroup gas and the other
(6000 km/sec component) with the intruder NGC7318b, occurred
$\sim 10^7$ years ago. This collision is perhaps part of the
larger scale collision which has resulted in the shock front
seen in the X-ray and radio continuum observations on the eastern side
of NGC7318b. The collision triggers an instantaneous starburst, 
giving rise to the strong MIR and H$_\alpha$ emission (its
$L_{15\mu}$ is about twice of that of the starburst in the collision
region of the Antennae Galaxies (Vigroux et al. 1996)).
The hot gas produced by the
collision has already cooled in the region of Source A presumably due
in part to the locally higher gas density which results in a  higher
cooling rate.

Interestingly, Source A sits also in a region where two long and 
faint optical arms seemingly run across with each other (Arp  \& Lorre 1976).
Both arms are likely being stretched out from NGC7318b, with the NGC7318a 
(an early type galaxy, Table 1) in the
background/foreground of the western arm (Moles et al. 1997).
It is tempting
to assume that these arms are products of tidal interactions between
NGC7318b and the other two galaxies (NGC7318a and NGC7319), and the
starburst in Source A is triggered by the interaction of these two
arms of NGC7318b. Strong evidence {\it against} this scenario,
and favoring the IGM--intruder collision scenario, is the
presence of a strong 6600km/sec component,
which is in fact the dominant component with apparent connection
with the shock front, in the H$_\alpha$ emission
in Source A (Fig.2). If there is no participation of the IGM
in the starburst, the 6600 km/sec component is not at all expected.
The fact that the H$_\alpha$ emission in the Source A region is dominated
by the 6600km/sec component conclusively demonstrate
that most of the star formation in Source A region must be
happening in the IGM.

In addition, the time scale for the development of 
tidal tails is more than a few 10$^8$ years (Barnes  \& Hernquist 1992;
Elmegreen et al. 1993)
while the crossing time of NGC7318b is only a few 10$^7$ years, too short
for the tidal effects to act. Indeed it is more likely that
the two optical `arms' are not `tidal arms' but are, together with Source A, 
products of some hydrodynamical processes induced by 
the high speed collision between the intruder and the rest of the group.
The eastern `arm' is almost certain to be associated with
the shock front (Moles et al. 1997)
although the mechanism for the western `arm' is still not very clear. 

Once the IGM environment of the starburst in Source A is determined,  
the next question is: could it be due to star formation in the
cooled gas in the aftermath of the high speed 
collision, similar to that proposed
for the star formation in `cooling flows' (Fabian et al. 1991)?
It has
been strongly argued, because of the red colors of the `cooling flow'
galaxies, that if star formation does occur in `cooling flows' it has
to be severely deficient in massive stars (Fabian et al. 1991).
This is contrary to what we are 
seeing in Source A where the high EW(H$_\alpha$)
clearly indicates a starburst with a lot of massive stars. Hence
the `star formation in cooling flow' mechanism is irrelevant here.
It is more likely that the starburst in Source A is triggered
by the gravitational instability casued by some hydrodynamical 
processes which in turn are induced by the
high velocity collision between the IGM and the intruder.
This might have some similatities, albeit in a much larger scale,
to the triggering mechanism of the 
so called `self-propagating' star formation mode (Blaauw 1964) 
in the Galactic disk, which is associated with
the shocks generated by HII regions (Elmegreen \& Lada 1977)
and by supernova remnants (Herbst \& Assousa 1977).

In summary, our conclusion that the starburst in the Source A region
is triggered by the high speed ($\sim 1000$ km/sec)
collision between the IGM and the intruder
NGC7318b is supported by the following facts:
\begin{enumerate}
\item The H$_\alpha$ data show that the star formation in Source A
is occurring both in the IGM (the 6600 km/sec component) and in
the intruder (the 5700 km/sec component). In particular,
the fact that the H$_\alpha$ emission in the Source A region is dominated
by the 6600km/sec component conclusively rules out any interpretation
involving only processes confined within the intruder NGC7318b.
\item The age of the starburst ($\sim 1$--- 2 $10^7$ yrs)
estimated from the MIR, H$_\alpha$ and NIR data is consistent with the
dynamical time scale of the high speed collision ($\sim 10^7$ yrs).
The probability that both the burst and the collision
are happening simultaneously (within such a short time scale) would be
very low if the former is not causally related to the latter.
\item The HI observations of Shostak et al. (1984) show that both
the intragroup gas and the gas associated with NGC7318b have local 
concentrations in the Source A region. These cold gas components are not
only the participants of the high speed collision which triggers the
starburst, but also the reservoirs of the
necessary material for the starburst to proceed.
\item The morphology of the 6600km/sec component of the H$_\alpha$
emission shows a possible connection between Source A and the shock
front, indicating that Source A is likely part of a larger scale
collision that is still going on between the IGM and the intruder.
\item Although Source A has no detectable X-ray emission, it is
surrounded by X-ray emitting hot gas, again indicating a
connection between the shock front (with which the X-ray
emitting hot gas is associated) and the starburst.
The high pressure imposed by the surrounding hot gas may
have also helped sustain the starburst by confining the cold gas in that
region.
\end{enumerate}

\section{Discussion}
\subsection{Theoretical implication} 
Theoretical works on interaction-induced starbursts are still in
their early stages (Mihos  \& Hernquist 1996).
Most of the studies have concentrated
on the effects of the gravitational tidal force,
while the effects of the
hydrodynamic shocks generated by the collision of two gas condensations
have been left out (Noguchi  \& Ishibashi 1986;  Mihos et al. 1993).
The scenario envisionized by Jog  \& Solomon (1992)
for starbursts induced by low velocity ($\sim 100$---300
km/sec) galaxy-galaxy collisions is difficult to apply to Source A. In that
scenario a starburst occurs when the
Giant Molecular Clouds (GMC's) preexisting in the two colliding disks
are compressed by the hot shocked gas. In the case of Source A,
it is not clear whether the IGM contains GMC's
(the CO survey of SQ by Yun et al. (1997)
unfortunately did not cover the Source A region).
Even if there are indeed preexisting GMC's in the Source A region, 
the colliding velocity is so high ($\sim 600$---1000 km/sec) and
close to the typical escape velocity in galaxies that the shocked gas
may be left behind when the colliding parties move away from each
other while GMC's are not affected by the collision because of the low
filling factor (Jog \& Solomon 1992). Clearly more theoretical
work, in particular simulations with a full treatment of gas
dynamics, is needed to understand what is going on in starbursts
such as Source A.

\subsection{A new mechanism for Butcher--Oemler effect?}
Events like Source A are very rare in the local universe.
However in the earlier universe, in particular in young clusters
where high velocity encounters involving late type galaxies are
frequent and a large part of the intracluster space may be filled
by cold stripped gas, collision between gas rich galaxies and
the cold intracluster medium (ICM) may be quite common. Whether
the starbursts so induced are at least partially responsible for
the enhanced star formation in those young, distant 
clusters (Butcher  \& Oemler 1984),
needs to be further explored
(for an alternative mechanism, see Moore et al. 1996).


\vskip2cm 
We are indebted to Jack Sulentic who helped in the acquisition of
H$_\alpha$ data. CX acknowledges very stimulating discussions with
Jack Sulentic, and constructive comments of George Helou,
Nanyao Lu, and Barry Madore. Part of the work is supported by 
NASA grant for ISO Data Analysis. NED is supported by NASA at IPAC.


\clearpage

\begin{deluxetable}{ccccccccc}
\tablewidth{0pt}
\tablecaption{ }
\tablehead{
& R.A. & Dec. & Type & $v_r$  & f$_{15\mu}$\tablenotemark{1} 
& H$_\alpha$\tablenotemark{2} & EW(H$_\alpha$)\tablenotemark{3} 
& K$'_{21}$\tablenotemark{4}  \nl
\cline{2-9} \nl
& \colhead{(J2000)}
& \colhead{(J2000)}
&
& \colhead{km/sec}
& \colhead{mJy}
& \colhead{10$^{-13}$erg/cm$^2$/sec}
& \colhead{\AA}
& \colhead{mag}
}
\startdata
NGC7319 & 22h36m03.5s & +33d58$'$33$''$ &
SBb (Sy2) & 6764 & 79.8 & 1.41$\pm 0.30$ & 7.6$\pm 2.5$  & 10.03 \nl
&&&&&&&& \nl
NGC7318a & 22h35m56.7s & +33d57$'$56$''$ &
Epec     & 6630 & & &     \nl
&&&&&&&& \nl
NGC7318b & 22h35m58.4s & +33d57$'$58$''$ &
SBbc     & 5774 &  &  &  \nl
&&&&&&&& \nl
NGC7318a/b\tablenotemark{5} 
&   & &  & &  19.3 & 1.08$\pm 0.16$ & 5.0$\pm 1.2$  & 9.23 \nl
&&&&&&&& \nl
NGC7317 & 22h35m52.0s &  +33d56$'$41$''$ &
E         & 6599 & 2.1  &   &   &  \nl
&&&&&&&& \nl
NGC7320 &  22h36m03.5s &  +33d56$'$54$''$ &
Sd  & 786 & 27.5 & & & 10.40 \nl
&&&&&&&& \nl
Source~A & 22h35m58.7s& +33d58$'$55$''$ &
SF region &  & 11.9 & 1.27$\pm 0.13$ & 115$\pm 15$ & 15.09 \nl
&&&&&&&& \nl
Source~B\tablenotemark{6}  & 22h36m10.2s &+33d57$'$21$''$ &
SF region &  & 1.8 &  0.14$\pm0.01$ & 85$\pm 10$ &  \nl
&&&&&&&& \nl
\enddata
\tablenotetext{1}{The 15$\mu m$ flux density, measured through 
aperture photometry covering areas included in 2$\sigma$ contours.
The uncertainties are about 30\%, dominated by the errors in the calibration.}
\tablenotetext{2}{H$_\alpha$ flux, corrected for the
[NII$\lambda$6583/$\lambda$6548] contamination. For NGC7319, the Sy2 galaxy, the ratio
[NII$\lambda$6583]/H$_\alpha =1.8$ is taken from Keel et
al. (1985). For other galaxies and star formation regions
the [NII$\lambda$6583]/H$_\alpha$ ratio is assumed to be 0.4 (Kennicutt
et al. 1994). The
errors are dominated by the uncertainties
in the continuum subtraction. No extinction correction is applied. 
}
\tablenotetext{3}{The equivalent line width of the H$_\alpha$ emission,
which is the ratio between the H$_\alpha$ flux (in erg/cm$^2$/sec)
and the flux density of the underlying continuum (in
erg/cm$^2$/sec/{\AA}). The continuum is estimated from the R-band flux.
For NGC7319 and NGC7318a/b the continuum includes the
emission from the entire galaxy and from the galaxy pair, respectively. 
For Source A, both the H$_\alpha$ flux and the continuum flux density
are measured through a rectangular aperture of $30''\times 25''$.}
\tablenotetext{4}{K$'$ ($2.1\mu m$)
magnitude measured through isophotal photometry down to K$'=21$ mag/acrsec$^2$.
NGC~7317 is not fully covered by the K$'$ image. Source B is not
detected in K$'$-band. The errors of the K$'$ magnitudes are less than
0.2 mag.}
\tablenotetext{5}{The fluxes of this pair of galaxies cannot
be separated reliably. Hence they are reported jointly for the pair.}
\tablenotetext{6}{Like Source A, although much fainter, 
Source B is also a star formation region in the IGM. On the other hand,
this source is far away from the shock front and apparently has
not been involved in the current collision between NGC7318b and the IGM.
Located at the end of one of the tidal tails pointing to 
NGC7320c (Arp \& Lorre 1974), presumably caused by 
{\bf previous} interactions between SQ
and NGC7320c (Moles et al. 1997), 
Source B is perhaps a star formation condensed out of 
the tidal tail (Mirabel et a. 1992).
}
\end{deluxetable}


\clearpage


\begin{figure}
\caption{
ISOCAM LW3 (15$\mu m$) image (false color) of Stephan's Quintet
field. The scale is in logarithm and the units are 0.1 mJy/pixel
(pixel$=6''\times 6''$). As shown by the chromatometer, the range of
the surface brightness shown in the figure is $-1 \leq \log 
[S_{15\mu m}/(0.1\; mJy/pixel)] \leq 2$.
The five members of the group, including the foreground galaxy
NGC7320, are clearly detected. Four other sources outside the galaxies
are marked in the image: Source A and Source B are star formation regions
associated with the group (see Table 1). 
Source C is a foreground star, and Source D is a background galaxy. 
\label{fig:lw3}
}
\end{figure}

\begin{figure}
\caption{
K$'$ (2.1$\mu m$) image overlaid by contours of the 
H$_\alpha$--[NII] emission, covering the same field of sky as in
Fig.1. The K$'$ image is in logarithmic scale. The two components of the
H$_\alpha$--[NII] emission are marked by different colors:
the contours of the 6600 km/sec component are
in red and those of the 5700 km/sec component in blue.
The two contour sets have the same levels, which are
0.3, 0.9, 2.7, 8.1, 24.3 $10^{-16}$ erg/cm$^2$/sec/pixel
(pixel$=0.533''\times 0.533''$).
The first contour level is chosen to hide most of the
residuals of the continuum subtraction, the most important
source of the errors of the H$_\alpha$--[NII] maps.
}
\end{figure}

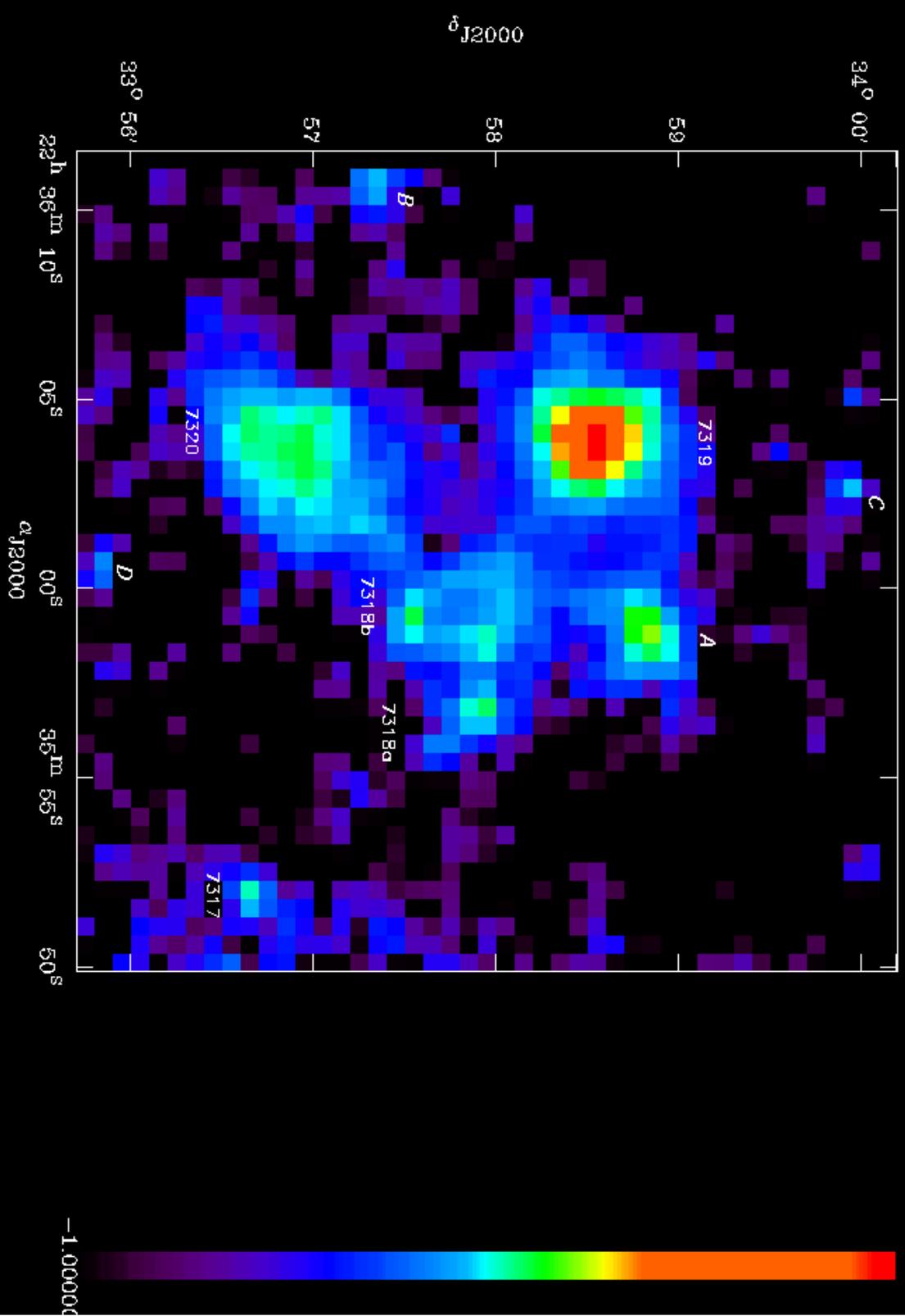
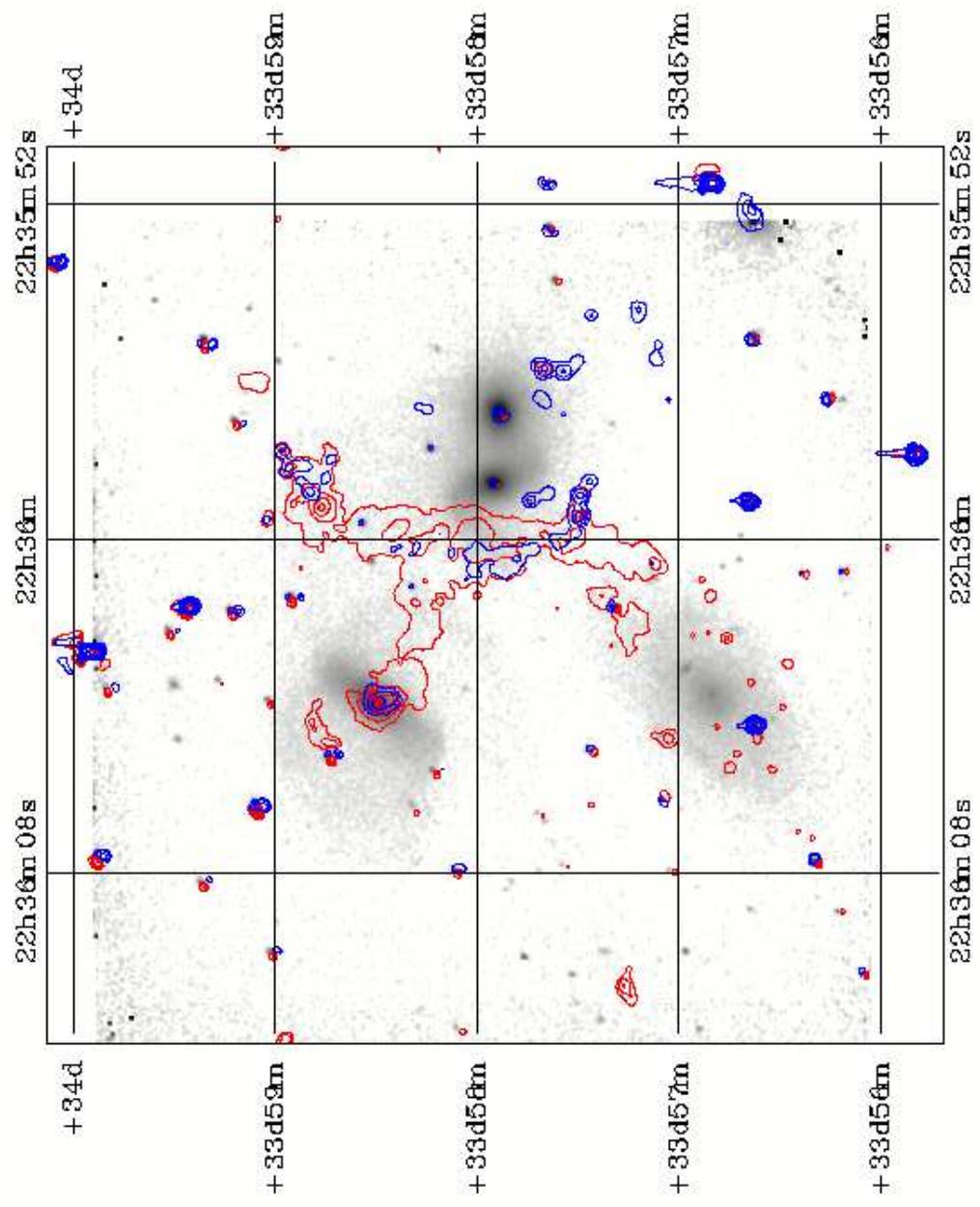
\end{document}